\documentclass[prl,preprint,showpacs]{revtex4}
\usepackage{amsfonts,amssymb,amsmath}
\usepackage{bm,epsfig}

\begin{document}
\title[Mesoscopic Ballistic Rectifiers]{Mesoscopic Rectifiers Based on
Ballistic Transport:\\Playing off Classical against Quantum Mechanics}
\author{Ragnar Fleischmann}
\email{ragnar@chaos.gwdg.de}
\author{Theo Geisel}
\email{geisel@chaos.gwdg.de}
\homepage{www.chaos.gwdg.de}
\affiliation{Max-Planck-Institut f\"ur Str\"omungsforschung und 
Fakult\"at Physik der Universit\"at G\"ottingen, Bunsenstr.\ 10,
D-37073 G\"ottingen, Germany\protect\footnote[1]{Permanent Address.}
\\and\\Institute for Theoretical Physics, University of
California, Santa Barbara, USA} 
\date{\today}

\begin{abstract}
Recent experiments on symmetry-broken mesoscopic semiconductor
structures~\protect\cite{song1,song2} have exhibited an amazing rectifying
effect in the transverse current-voltage characteristics with promising
prospects for future applications.  We present a simple microscopic model,
which takes into account the energy dependence of current-carrying modes
and explains the rectifying effect by an interplay of fully quantized and
quasi-classical transport channels in the system. It also suggests the
design of a ballistic rectifier with an optimized rectifying signal and
predicts voltage oscillations which may provide an experimental test for
the mechanism considered here.
\end{abstract}
\maketitle

The decreasing size of nanofabricated structures opens up new possibilities
for mesoscopic semiconductor devices by exploiting ballistic transport in
the 2-dimensional electron gas and quantization of confined electrons.
Details of the geometry are crucial for the functioning of such devices. In
the geometry of Fig.~1a, which was used in an experiment by Song et
al.~\cite{song1}, the symmetry was broken on purpose by introducing a
triangle as shown. When current is injected at the source (S) and drawn out
at the drain (D), a majority of charge carriers is deflected towards the
lower voltage probe (L). Naively thinking and in the spirit of the Hall
effect one might expect that a voltage difference builds up between the
upper (U) and lower (L) voltage probe. Since the sample is symmetric with
respect to the exchange of source and drain, the same voltage difference
would arise on reversing the current and the sample would work as a
rectifier.
 
\begin{figure}
\epsfig{file=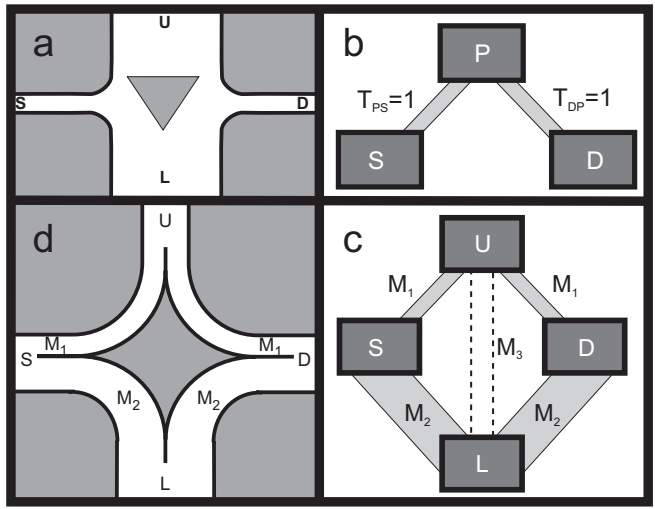, width=8.6cm}
\caption{(a) Geometry of the experimental setup used in
Ref.~\cite{song1}. The leads at $S$ and $D$ are approx.\ $400nm$ wide,
corresponding to approx.\ $20$ to $22$ modes at the equilibrium Fermi
energy (about $18meV$). The leads at $U$ and $L$ are approx. $3.2\mu m$
wide, corresponding to about $180-200$ modes. (b) A scheme of a voltage
probe. (c) A rectifier consisting of a combination of two such probes with
two or three channels. (d) Suggested geometry of an optimized ballistic
rectifier.}
\end{figure}

On second thought, however, our physical understanding of mesoscopic
systems leads us not to expect any voltage drop from top to bottom at
all. As will be argued in detail below, symmetry considerations and the
Landauer-B\"uttiker formalism~\cite{LBF} in its most common linear form do
not allow for such a voltage drop. It thus came as a surprise, when the
experiment by Song et al.~\cite{song1} revealed a rectifying effect.
An interpretation of the experimental result in Refs.~\cite{song1,song2}  
was based on the assumption of dissipation within the sample leading to
self-consistent electric fields and a current dependence of the
transmission probabilities through the sample. A phenomenological ansatz
was made, as this current dependence would be exceedingly complicated to
calculate in a microscopic model.

\begin{figure}
\epsfig{file=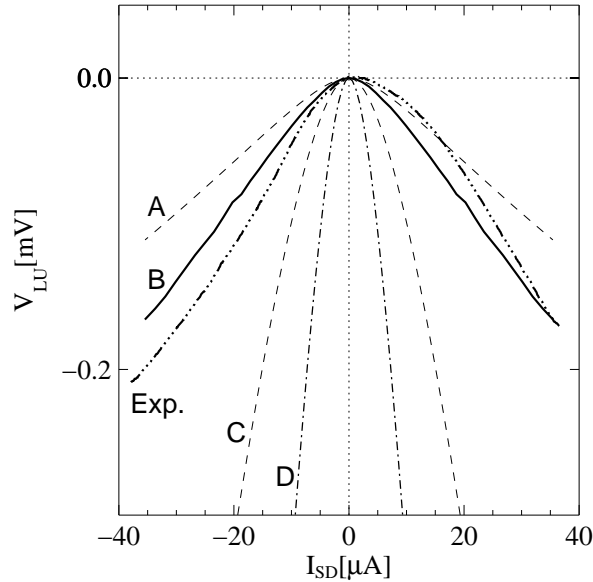, width=8.6cm}
\caption{(A-D): Transverse current-voltage characteristics for various
models of a ballistic rectifier as described in the text calculated for
temperature $T=4K$ and Fermi energy $\mu_F=18 meV$.  (Exp.): Experimental
curve after Ref.~\protect\cite{song1} shown for comparison.}
\end{figure}

In the present letter we present a microscopic explanation of the
rectifying effect which explicitely takes into account the energy
dependence of the number of transverse modes for a system of two
voltage probes. It does not require the existence of dissipation
in the leads, instead  the rectifying effect here originates in
the interplay between purely quantized and quasi-classical
transport in different channels of the system, which exhibit
different energy dependences. A calculation of the tranverse
current-voltage characteristics involving this mechanism shows
good agreement with the experimental observations. This mechanism
also leads to a prediction for the design of a ballistic rectifier
with an optimized rectifying signal. For strong currents it
predicts a reversal and even oscillations of the transverse
voltage, which may provide a test for the explanation presented
here.

Transport in mesoscopic systems like the one in Fig.~1a is
typically described by the linear Landauer-B\"uttiker-formalism
\cite{LBF}
\begin{equation}
I_i=\frac{2e}{h}\left[(M_i-R_i)\mu_i-\sum_{j\ne i}\overline{T}_{ij}\,\mu_j\right].
\label{eq:lb1}
\end{equation}
Here $I_i$ is the net current in lead $i$ connecting the sample to
a reservoir (contact) with chemical potential $\mu_i$. The leads are
assumed to be ideal quantum leads with $M_i$ modes. $R_i=\overline{T}_{ii}$
is the reflection coefficient, which describes back-scattering from the
sample into lead $i$ and $\overline{T}_{ij}$ are the transmission
coefficients form lead $j$ into lead $i$. Transport across the sample
is assumed to be purely elastic and dissipation and equilibration only
take place in the reservoirs.

A prominent result of the Landauer-B\"uttiker formalism is the
\emph{reciprocity relation} 
\begin{equation}
R_{ij,kl}(B)=R_{kl,ij}(-B).\label{eq:sym1}
\end{equation}
 Here $R_{ij,kl}=V_{kl}/I_{ij}$ is the resistance obtained by dividing the
voltage $V_{kl}$ measured between contact $k$ and $l$ by the current
$I_{ij}$ flowing from contact $j$ to $i$. For the system of Fig.~1a at zero
magnetic field we thus have
\begin{equation}
 R_{UL,SD}=R_{SD,UL}.\label{eq:sym2} 
\end{equation} 
  Because of the symmetry of the system there can be no voltage build-up
between $S$ and $D$ if the current is flowing from $L$ to $U$. Thus
$R_{UL,SD}=0$ and hence by means of Eq.~(\ref{eq:sym2}) we would expect
$V_{UL}$ to vanish identically -- in contrast to the experimental findings.
(We would of course obtain the same results by solving the
Eqs.~(\ref{eq:lb1}) directly.) To overcome this apparent contradiction Song
et al.\ \cite{song1,song2} suggested to include a phenomenological current
dependence of the transmission coefficients due to dissipation inside the
sample. In our treatment the necessary nonlinearity arises in the transport
equations, if we allow for varying numbers of modes in the leads. We will
show below that one needs not give up the conceptually attractive
assumption of purely elastic transport inside the sample, that was so very
successful in describing a wide range of experiments on transport in
mesoscopic system (for reviews see
e.g. \cite{mesotrans1,mesotrans2,mesotrans3}).

To achieve this aim we use the Landauer-B\"uttiker formalism in a more
general form~\cite{datta}. The current per unit energy (a quantity, which
we will call \emph{current density} in the following for simplicity)
injected into the sample from reservoir $i$ through lead $i$ at energy $E$ is
\begin{equation}
i^{+}_{i}(E,\mu_i,T)=\frac{2e}{h}M_i(E) f_i(E).\label{eq:im1}
\end{equation}
Here $f_i(E)=f(E,\mu_i,T)$ is the Fermi-distribution in reservoir $i$ at
temperature $T$. The outgoing current density in lead $i$ is
\begin{equation}
i^{-}_{i}(E,\{\mu_l\},T)=\frac{2e}{h}\left[ R_i(E) f_i(E)+\sum_{i\ne
j}\overline{T}_{ij}(E)f_j(E)\right].\label{eq:ip1}
\end{equation}
If we assume, that the transmission probabilities $T_{ij}$ are independent
of energy and the mode number, we can write
$\overline{T}_{ij}(E)=T_{ij}\,M_j(E)$ and Eq.~(\ref{eq:ip1}) becomes
\begin{equation}
i^{-}_{i}(E,\{\mu_l\},T)=\sum_{j}{T}_{ij}\,i^{+}_{j}(E,\mu_j,T)
.\label{eq:ip2}
\end{equation}
The incoming and outgoing currents are, respectively, 
\begin{equation}
I^{\pm}_i=\int_{\mu_0}^{\infty}i^\pm_i(E)\,dE,\label{eq:curr}
\end{equation}
where $\mu_0$ is an auxiliary quantity that is small enough so that
$f(\mu_0,\mu_i,T)\approx 1$ holds for all $i$, but is otherwise arbitrary and
will not show up in any measurable quantity.  The balance equation for the
current source with net current $I$ then is $I^{+}_{S}-I^{-}_{S}=I$ and for
the drain $I^{+}_{D}-I^{-}_{D}=-I$.  A voltage probe is characterized by
zero net current, i.e. $I^{+}_i=I^{-}_i$.

The number of modes $M_j(E)$ can be written as 
\begin{equation}
M_j(E)=\sum_n \Theta(E-\varepsilon_{j,n}),
\end{equation}
where the $\varepsilon_{j,n}$ are the energy eigenvalues of the transverse
modes in lead $j$ and $\Theta(x)$ denotes the Heaviside step function. For
leads with a hard wall (i.e. box-like) cross section of width $W_j$
and for electrons of effective mass $m^*$ we have
\begin{equation}
\varepsilon_{j,n}=\frac{(\hbar\pi n)^2}{2m^*W_j^2}.
\end{equation}
In this case the number of modes  can also be expressed as 
\begin{equation}
M(E)=\text{Int}\left[\frac{W_j}{\lambda(E)/2}\right]=
	\text{Int}\left[\frac{W_j\sqrt{2m^*E}}{\hbar\pi}\right],
\label{eq:mebx}
\end{equation}
where $\text{Int}[\,]$ denotes the integer part and
$\lambda(E)=h/\sqrt{2m^*E}$ is the de Broglie-wavelength of the electron at
energy $E$.  The incoming currents can be expressed as
\begin{equation}
I^+_j(\mu_j,T)=\frac{2e}{h}\sum_n\,\left\{kT
\ln\left[e^{(\varepsilon_{j,n}-\mu_j)/kT}+1
\right]+\mu_j-\varepsilon_{j,n}\right\}, 
\end{equation}
and the outgoing currents accordingly by integrating Eq.~(\ref{eq:ip2}).
These formulas were used in obtaining the numerical results presented below.

Let us now examine the very simple setup of Fig.~1b. A voltage probe
reservoir (P) is connected via two identical ideal leads to source and
drain. To simplify the calculations let us assume zero temperature, i.e.\
$f(E,\mu_i,0)=\Theta(\mu_i-E)$. Let us further assume
$\mu_S\ge\mu_P\ge\mu_D\ge\mu_0$, where $\mu_S$ and $\mu_D$ are given
and $\mu_P$ is to be determined, and let us distinguish the cases of narrow
and wide leads.  

\textbf{(a)} If $M(E)=M=\,$constant in the range between $\mu_0$ and
$\mu_S$ (this corresponds to narrow leads), the outgoing and incoming
current densities from and to the probe reservoir are simply given by:
$i^{+}_{P}(E)=2(2e/h)M\Theta(\mu_P-E)$ (the factor 2 arises because two
leads are connected to the same reservoir) and
$i^{-}_P(E)=(2e/h)\left[\Theta(\mu_S-E)+\Theta(\mu_D-E)\right]$. These are
trivially integrated and the current balance of the voltage probe
$I^+_P=I^-_P$ reads
\begin{equation}
\frac{4e}{h} M (\mu_P-\mu_0)=\frac{2e}{h} M (\mu_S+\mu_D-2\mu_0)
\end{equation}
and thus the chemical potential $\mu_P$ of the voltage probe is
independent of M and given by
\begin{equation}
\mu_P=(\mu_S+\mu_D)/2.
\label{eq:muqu}
\end{equation} 

\textbf{(b)} If the leads are assumed to be wide compared to the Fermi
wavelength $\lambda_F$, the number of modes will increase even under small
changes in energy. In the case of a hard wall channel of width
$W>>\lambda_F$ we can approximate Eq.~(\ref{eq:mebx}) by a smooth function,
i.e.
\begin{equation}
M(E)=Int\left[\frac{W}{\lambda(E)/2}\right]\approx
\frac{\sqrt{2m^*}W}{\pi \hbar}\sqrt{E}.
\label{eq:qcap}
\end{equation}
Since $M(E)$ grows as $\sqrt{E}$, the channels described in this
approximation may be called \emph{quasi-classical}, because the classical
energy surface likewise increases as the square-root of the energy.  The
current densities are $i^{+}_P(E)=2Q\sqrt{E}\Theta(\mu_P-E)$ and
$i^-_P=Q\sqrt{E}[\Theta(\mu_S-E)+\Theta(\mu_D-E)]$ with
$Q=\sqrt{16m}eW/h^2$. The current balance now reads
\begin{equation}
\frac{4Q}{3}\left(\mu_P^{3/2}-\mu_0^{3/2}\right)=
\frac{2Q}{3}\left(\mu_S^{3/2}+\mu_D^{3/2}-2\mu_0^{3/2}\right),
\end{equation}
and we find
\begin{equation}
\mu_P=\sqrt[3]{\left(\mu_S^{3/2}+\mu_D^{3/2}\right)/2}^2>(\mu_S+\mu_D)/2,
\label{eq:mucl}
\end{equation}
i.e.\ the potential $\mu_P$ deviates from the mean Eq.~(\ref{eq:muqu}). As
illustrated in Fig.~3 this is because the net current is transported from
$S$ to $P$ by a larger number of modes in the energy interval
$[\mu_P,\mu_S]$ than from $P$ to $D$ in the energy interval $[\mu_D,\mu_P]$
and thus $\mu_P$ raises above the mean to compensate for the additional
current.

\begin{figure}
\epsfig{file=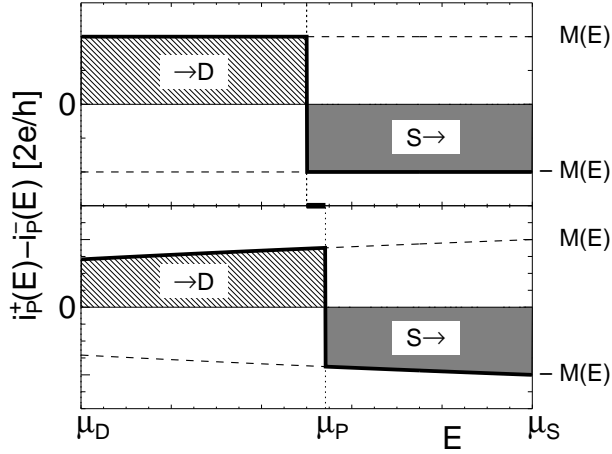, width=6.0cm}
\caption{Current balance in the voltage probe P and adjustment of the
chemical potential $\mu_P$ at zero temperature. Current enters P from S in
the energy interval $[\mu_P,\mu_S]$ and exits from P to D in the interval
$[\mu_D,\mu_P]$. The total current $I^+_P-I^-_P$ is determined by the
shaded areas and must balance to zero. (a) For narrow leads ($M(E)=$const.)
this obviously determines $\mu_P$ as the mean $(\mu_D+\mu_S)/2$. (b) In the
quasi-classical case $\mu_p$ must shift to higher values to counterbalance
the increase of $M(E)$ with energy.}
\end{figure}

Equations~(\ref{eq:muqu}) and (\ref{eq:mucl}) suggest how to create a
rectifier based on ballistic transport: As sketched in Fig.~1c let us
consider two separate pathways from source to drain, each via a voltage
probe as in the above example (ignoring the third dotted pathway for the
moment). If on both paths the number of modes is constant, the voltage
probes will each be at the mean chemical potential between source and
drain, i.e. there will be no voltage drop from top to bottom (this is what
we showed based on the reciprocity relation in the beginning). The same
holds true, if the current density grows identically with energy in both
channels, e.g. for two quasi-classical channels. If on the other hand one
path is narrow, i.e. has constant mode number $M_1$, whereas the other is
wide with an increasing number of modes $M_2(E)$, we can observe a voltage
drop between the probes, as indicated by the interval marked by the fat
line in Fig.~3. Reverting the current yields the same voltage drop due to
symmetry and thus we achieve the rectification of the signal.

In order to explain the experimental results of Ref.~\cite{song1} we need
to assume a third channel connecting the probes U and L in Fig.~1c (dotted
lines) with $M_3(E)$ modes, since there is a direct connection between U and
L in Fig.~1a. Now we consider finite temperatures again and choose $M_1=1=$
const.\ in the energy window, whereas for $M_2(E)$ and $M_3(E)$ we use the
quasi-classical Eq.~(\ref{eq:qcap}) with $M_2(E_F)=20$ and
$M_3(E_F)=15$. These mode numbers were estimated by classical numerical
calculations of the transmission coefficients $\overline{T}_{US}$ and
$\overline{T}_{LS}$ in the geometry of Fig.~1a modeling the
experiment. Curve A in Fig.~2 shows the resulting rectifying signal of this
setup. If we give up the quasi-classical approximation for the wide
channels and use the explicit sums instead, we obtain curve B, which nicely
agrees with the experimental result (note that the experimental curve is
not entirely symmetric due to an unintentional slight asymmetry of the
sample about the vertical axis).

Based on the above mechanism we can now suggest the design of a ballistic
rectifier with an optimized rectifying signal. If one manages to suppress
the third channel between U and L, the voltage drop $V_{UL}$ can be
enhanced. This is shown by curve C of Fig.~2 for the model of Fig.~1c
without the dotted channel and using the quasi-classical approximation for
the wide channel. Note that this curve can easily be calculated
analytically for T=0.  Using Eq.~(\protect\ref{eq:muqu}) and
(\protect\ref{eq:mucl}) one obtains $\mu_U$ and $\mu_L$, and hence the
voltage difference, as a function of $\mu_S$, while leaving $\mu_D$
constant. The total current $I$ is given by the sum of the individual
currents from the source S to U and to L, i.e.\
$I=2Q/3\,(\mu_S^{3/2}-\mu_L^{3/2})+2e/h\,(\mu_S-\mu_U)$, which is the
current-voltage characteristics in analytic form.  If again we give up the
quasi-classical approximation for the wide leads, we find an even stronger
rectifying signal as shown in curve D. As a realization of such a system
which suppresses the third channel we suggest the geometry shown in
Fig.~1d.

An interesting phenomenon arises in these structures, which combine narrow
and wide leads, when the narrow channel opens up a new mode within the
energy window: the voltage $V_{UL}$ undergoes a change in sign! This is
demonstrated in Fig.~4, where the dotted line again shows curve D from
Fig.~2. The solid line corresponds to slightly wider leads in the upper
channel whereby $\mu_S$ becomes larger than $\varepsilon_2$ of the narrow
leads. This curve will eventually turn back to negative voltages as $I$
(following $\mu_S$) increases and shoot up again, when the third mode opens
in the upper channel. This change in sign or even oscillations should be
observable in an appropriate experiment, which would provide a test for the
mechanism of ballistic rectifiers presented in this paper.

\begin{figure}
\epsfig{file=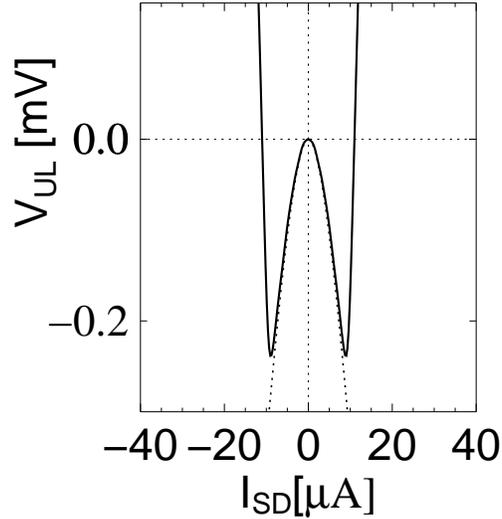, width=6.0cm}
\caption{The voltage $V_{UL}$ undergoes a change in sign as a second mode
opens up in the narrow channel of Fig.~1c.}
\end{figure}

\begin{acknowledgments}
The authors would like to thank Roland Ketzmerick for stimulating
discussions.
This research was supported in part by the the National Science Foundation
under Grant No. PHY99-07949.
\end{acknowledgments}

\end{document}